\documentclass[11pt]{article}
\usepackage{fullpage}
\usepackage{setspace}
\usepackage{graphicx}
\usepackage{times}
\usepackage{enumerate}
\usepackage{color}
\usepackage[pdftex]{hyperref}

\begin{document}

\title{GalViz: Visualize Social Bookmark Management} 
\title{Context Visualization for Social Bookmark Management}
\author{Lilian Weng and Filippo Menczer\\
School of Informatics and Computing, Indiana University Bloomington, USA
}
\date{}

\maketitle

\begin{abstract}
We present the design of a new social bookmark manager, named \textit{GalViz}, as part of the interface of the GiveA\-Link system. Unlike the interfaces of traditional social tagging tools, which usually display information in a list view, \textit{GalViz} visualizes tags, resources, social links, and social context in an interactive network, combined with the tag cloud. Evaluations through a scenario case study and log analysis provide evidence of the effectiveness of our design.\\
\\
\textbf{Keywords}: social tagging, social links, Web content, visualization, interface, layout, network, GiveALink
\end{abstract}


\section{Introduction}
Social tagging is a flexible and powerful way to assist users who collaboratively organize and share Web resources, as is widely done in many popular Web~2.0 applications. Various kinds of online resources can be tagged, reflecting user interests and making such resources more easily accessible to the public. 
The basic relationship in social tagging is defined by the \textit{triple}, a tripartite combination of a user, a tag and a resource. To clearly display and manage a large number of triples is a pretty hard problem, since they involve complicated relationships among objects of three different types. 

Current social bookmark managers display collected resources in simple list views, distributed across multiple pages. Such a linear display of resources neglects contextual information and weakens the relationships between resources and tags.  Socio-semantic connections among users through shared annotations driven by common interests are hidden, since most existing systems completely separate the personal and global views. Semantic relationships among related tags and among related resources are deemphasized as well. Our design aims to ameliorate these problems for facilitating the management of complex relationships hidden in the social annotations. In particular, the design adds value to the user experience by visualizing tripartite relationships as networks.

\section{Motivation}
Social tagging collections, also known as social bookmarks or \emph{folksonomies}, are generated by spontaneous behavior of users, helping people manage and organize their resources~\cite{millen2005}. 
%
Existing social tagging sites usually incorporate components such as resource lists assisted with tag lists, tag clouds and hierarchical clusterings. However, each display has deficiency limiting its functionality and usability. The imperfection of current systems encourages us to design and develop a better alternative application.

\begin{figure*}
  \centering
  \includegraphics[width=\textwidth]{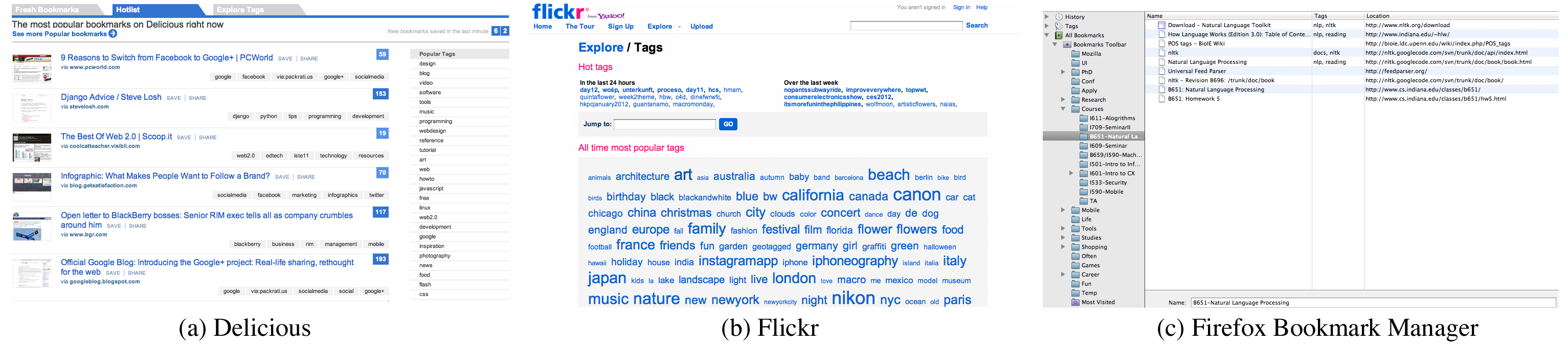}
  \caption{Interfaces of existing social bookmark managers: (a) Delicious.com (b) Flickr (c) Firefox Bookmark Manager.}
  \label{fig:oldinterface}
  \vspace{-1em}
\end{figure*}

\textbf{List View:} 
List views are widely adopted for displaying a group of items of the same type. The interface of \textit{Delicious}, as an example, consists of a list of resources and a list of all tags (Fig.~\ref{fig:oldinterface}a). The resource list occupies most space of the page, emphasized as the main content. Tags are crowd-sourced keywords for indexing resources, so the tag list is treated as an auxiliary sidebar. 
In this case, resources and tripartite relationships are displayed linearly. List views sequentially present the resources according to a certain order, hiding the links among resources, other similar resources and semantically related tags. Key contextual information, such as similar sites, people who have interests in it, and their collections, could be very valuable for fostering a better understanding of the site and the exploration of other relevant and interesting objects; but it is missing.

\textbf{Tag Cloud:} 
The tag cloud is a visual depiction of tags in the whole collection to facilitate tag browsing (Fig.~\ref{fig:oldinterface}b). Tags are arranged in alphabetical order; font sizes and colors are configured to highlight the most popular ones. Tag clouds provide users with a global view to learn which tags are used by most people, but they possess several disadvantages too.
Flat tag clouds are not sufficient to provide a semantic, rich and multidimensional browsing experience over large tagging spaces, due to their inefficient navigations, downgrading of alternative views, and lack of emphasis on the semantic relationships~\cite{hearst2008, Quitarelli2007facetag}. Tag clouds obscure useful information by having a strong bias towards very popular tags~\cite{deutsch2011, gupta2011overview}. 
They usually don't have enough space to contain all the tags, so a portion of resources still remain inaccessible through the tag clouds~\cite{gupta2011overview}. 

\textbf{Browser Bookmarks:} 
The hierarchical display of tags is naturally embedded in bookmark managers of Web browsers (Fig.~\ref{fig:oldinterface}c). The browser tagging system is designed to be very similar to a file explorer, where tags are organized in a folder-like hierarchical structure. 
Resource categorization is restrict\-ed by the structure of tags, and only the child-parent relations get apparent. Therefore, when the collections of bookmarks become larger, it is hard for users to efficiently find a specific resource or discover related items~\cite{passin2003}. Without support for cross references, each resource instance can only belong to one folder. Although multiple copies of the same resource can be assigned to different tags, they are distinct instances, leading to duplication and inconsistency. 


\section{Related Work}

There is sparse literature on the design principles of social tagging tools~\cite{deutsch2011, farooq2007, shiri2009}. Some work has been done on building hierarchy structure from the social tagging to uncover the hidden child-parent semantics~\cite{begelman2006tagClustering, shepitsen2008recomm}. 
%
Visualization, as a powerful type of social bookmarking tool, has been studied and utilized in several existing designs.
Visualization of hyperlinks between Web pages was adopted to enhance adaptive navigation~\cite{Tomsa2008}.
Cluster Map, a social bookmark visualization tool, highlighted the relationships among users and bookmarks to identify tag and community structures~\cite{klerkx2009}. Unlike ClusterMap, \textit{GalViz} is designed to emphasize the semantic relationships between tags and resources to help users manage existing resources and discover new ones.
Besides, the graphical interface was shown to be useful for distributed collaborations and interactions on social bookmarking~\cite{collomb2010}.
Graphical visualization of concept networks was integrated into several Web applications as an innovative interactive user interface~\cite{infomous, visuwords}. Most of such existing applications display context of items of the same type, but \textit{GalViz} is able to provide the visualization of heterogenous networks among two different objects, tags and resources.

\url{GiveALink.org}, as a research-oriented social tagging platform, broadly examines several aspects of social tagging to foster the construction and applications of socially driven semantic annotation networks. Previous research includes the design of effective similarity relationships~\cite{ben09}, social spam detection~\cite{ben09_2}, and social tagging games as incentive for collecting high-quality annotations~\cite{ weng2011cgames}. Former work in GiveA\-Link on exploratory navigation interfaces~\cite{Donaldson08ht} and bookmark management~\cite{sigir09_GaL_demo} have significant influence on the new design presented in the paper.

\section{Design Goals}

\begin{figure*}[t]
\centering
\includegraphics[width=0.9\textwidth]{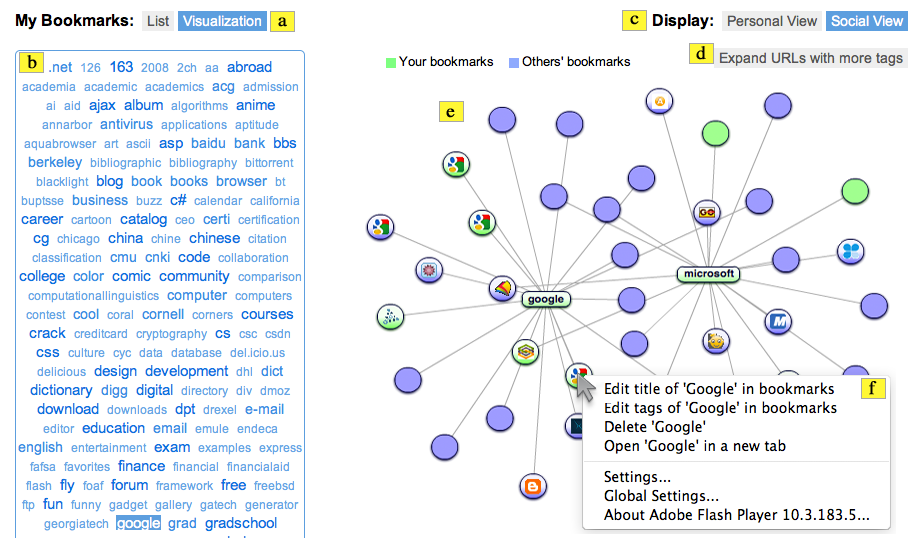}
\caption{Implementation of \emph{GalViz}:
(a) Choose from two modes, either the traditional list view or the network visualization. 
(b) Tag cloud of all personal tags; the font size of a tag is proportional to its popularity.
(\/c) A switch between the personal view and the social view. The social view displays additional relevant resources and tags based on other people's collections. The local items are green nodes, and the global items are purple ones. 
(d)~The filter allows one to expand available resources in the network with more related tags. 
(e) Visualization of tags and resources. 
(f) Contextual menu of a node. The sample menu is for a local resource, so the user has full control over it, including editing the title, changing the tags, or removing it. If the resource doesn't belong to the user, she has to add it to her local collection prior to making further changes.}
\label{fig:implementations}
\vspace{-1em}
\end{figure*}

\emph{GalViz} is an innovative bookmark management tool in the \emph{GiveALink} system (Fig.~\ref{fig:implementations}). We aimed to achieve several design goals, to help overcome the weakness of existing systems and improve user experience. Network display, contextual maps, and social relationships are three core concepts.

\subsection{Facilitate Annotation Management with Visualization}
\emph{GalViz} visualizes personal bookmarks in an interactive network. Each node represents a Web resource or a tag; related objects are connected. The default network is centered at selected tags, and gets expanded by relevant triples. It is easier to interpret than pure text~\cite{chi2002}. Furthermore, 
users only need to concentrate on a small area of the visualization, and therefore they can quickly spot the target node by the text or favicon.
%
The network provides an interactive way to easily edit or remove any node through a contextual menu. The menu shows different options according to the characteristics of the node: whether it is a resource or a tag, and whether it belongs to the user's personal collection. Dragging and dropping nodes create operation shortcuts. 
A \textit{GalViz bookmarklet} allows users to easily add new resources and view their semantic context.

\subsection{Provide Contextual Maps}
Currently, neither browsers nor popular social tagging sites visualizes a page's semantic context beyond its content.
Therefore, when users need to tell what a page is about, they learn based on the information inside the page itself. To improve this, \textit{GalViz} strives to provide valued and helpful knowledge by exposing links with correlated tags and resources.
Users are therefore provided with contextual maps of semantic relationships of a tag or Web page in terms of related tags and pages. Users can use such a map as an alternative channel to understand the content and context of the target, as well as to navigate the Web by visiting or exploring related objects.

\subsection{Highlight Social Relationships}
In social tagging systems, no matter whether users communicate directly, they are always potentially connected by shared interests in certain resources or similar agreements on the resources' descriptions. One limitation of the current interface~\cite{delicious} is that personal resources and the global repository are completely separated, and consequently users cannot easily discover their social connections with others. People have different views of same resources, so learning others' annotations can help improve one's understanding and optimize one's collection. \textit{GalViz} brings the idea of \textit{social view} instead of \textit{global view}. The social view expands the local (personal) resources and tags of a user with globally related triples, making it easier to find out what other people view as similar objects.

\subsection{Do Not Overwhelm Users with Information}
Information overload is a common problem faced by many crowdsourcing sites. Once a large group of users get into the system and contribute to the knowledge repository, it is impossible for anyone to browse every piece of information. Therefore, customization of the results becomes necessary. To avoid overwhelming users, we propose a set of content filters in the design to limit the network expansion, and to switch between personal and social view.

\section{Implementation}

\emph{GalViz} incorporates two styles of browsing, the \emph{list} mode and the \emph{visualization} mode, as well as a \emph{bookmarklet} (Fig.~\ref{fig:casestudy}(a)) gadget helping add new resources.

\textbf{Visualization}:
The visualization of bookmarks as a network with rich contextual information is the core design of \textit{GalViz}. As Fig.~\ref{fig:implementations} shows, the visualization mode is formed of a tag cloud on the left and a main bookmark network canvas on the right. Several options on the top endow users with the ability to switch between different views or filter out the results.

\textbf{List}:
We provide a traditional list view to assist users gradually accommodate to the new interface. The list view is also used for evaluation by comparison between the two modes. Similar to the visualization mode, the list interface has a tag cloud on the left, and the main content on the right where resources of selected tags are listed. Personal and social view separate the local resources and globally popular ones, with a switch similar to Fig.~\ref{fig:implementations}(a).

\textbf{Bookmarklet}:
The bookmarklet is a small applet, working as a normal bookmark item but with customized javascript code to provide additional functions beyond simply opening a Web page. The \textit{GalViz} bookmarklet is an extension of the visualization service providing easy access to the contextual map of the current Web page.

\section{Case Study}

\begin{figure}
\centering
\includegraphics[width=0.4835\columnwidth]{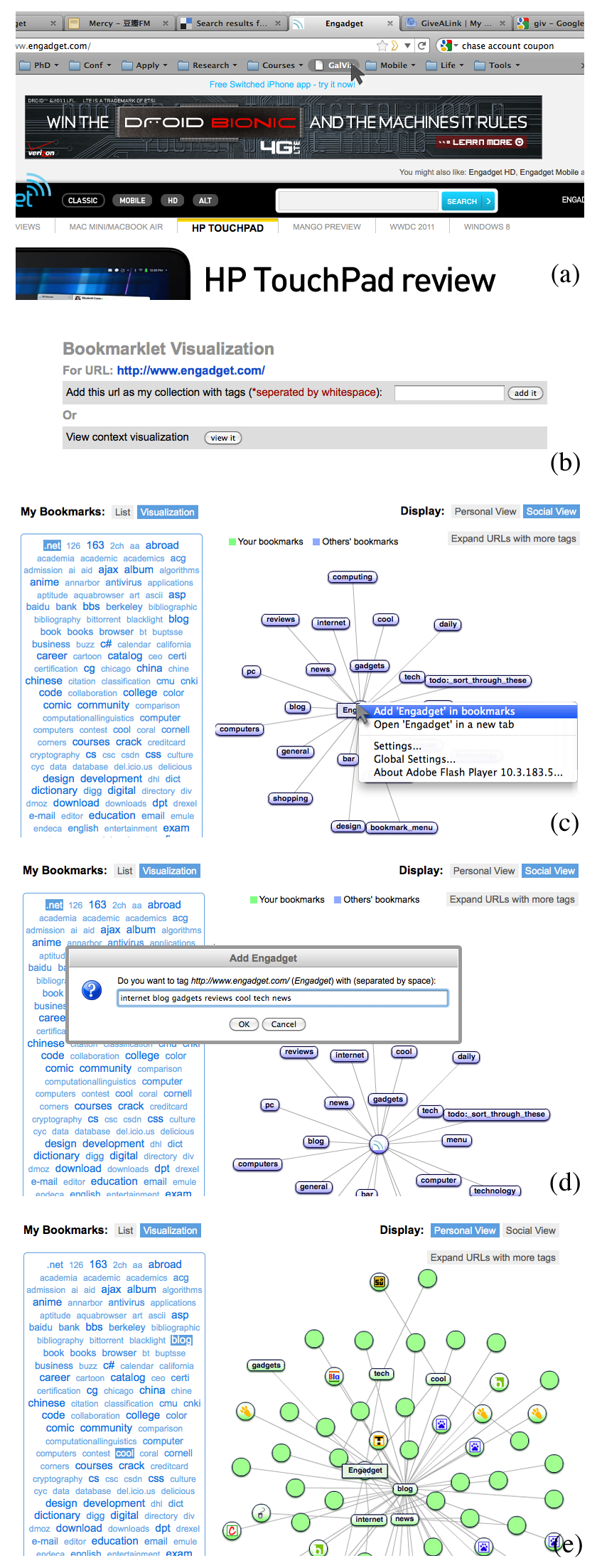}
\vspace{-1em}
\caption{Find out the context of the site \url{Engadget.com} and add it into the  personal collection through \textit{GalViz}: (a) Use \textit{GalViz} bookmarklet; (b) Choose to see the context; (c) Add it into the personal collection; (d) Customize recommended tags; (e) Check the final visualization.}
\label{fig:casestudy}

\end{figure}

\begin{description}

\item [Scenario:] A friend recommends the site \url{Engadget.com}.

\item [Goal:] To learn the content and context of \emph{Engadget} through the social tagging system, and then annotate it with a set of proper tags in the personal bookmark collection.

\item [Solution:] Let us consider how a user accomplishes the task through \emph{Delicious} and \textit{GalViz,} respectively:

(1) \textit{Through Delicious}: First, the user visits \textit{Delicious} and searches for ``engadget.com''. She should be able to see top popular tags about \textit{Engadget}. Second, the user clicks on ``Save this site" on the top right corner and fills in the pop-up dialog box to tag the bookmark.

(2) \textit{Through GalViz}: While browsing the \textit{Engadget} site, the user can access \textit{GalViz} by simply clicking the \textit{GalViz} bookmarklet (Fig.~\ref{fig:casestudy}a). After choosing ``View context visualization'' (Fig~\ref{fig:casestudy}b), he can see the contextual information about \textit{Engadget}. Through the right-click menu, the target site, as the center of the visualization, is added into the personal view (Fig.~\ref{fig:casestudy}c) with tags customized on the system recommendations (Fig.~\ref{fig:casestudy}d). The final visualization will be the personal view of \textit{Engadget} associated with user tags and other relevant Web resources (Fig.~\ref{fig:casestudy}e).

Users need to type in the required information and read the text in detail to look for proper links in \emph{Delicious}. \textit{GalViz} provides a more transparent and embedded service through the bookmarklet, and visualizes the contextual information in a simple but insightful way. 

\end{description}

\section{Preliminary Log Study}

During an evaluation period, we assigned \emph{GiveALink} users with equal chances to see either the list or the visualization mode when they navigated to the bookmark manager. Users are allowed to do any task they attend to do. Our log recorded 38,181 actions by 6,310 users. 
Analysis of the log reveals several advantages of the visualization interface. 
For each mode, a series of continuous clicks is defined as a \textit{session}. As shown in Table~\ref{table:log}, the visualization can \textbf{attract more usage}, as there are more sessions in the visualization mode (1,192) than in the list view (960). 
Each session in the visualization mode lasts 172.3 seconds on average, 25 seconds shorter than the average session length of the list. With the help of the interactive network visualization, users are shown to interact more with the system during a shorter time period. 
With a reasonable assumption that users start using the \emph{GiveAlink} bookmark manager with a certain goal and leave once the goal is achieved, we can infer that the visualization mode can \textbf{facilitate user tasks more efficiently}. 
Clicks involving tag selection, resource selection and content editing are considered as content-related actions. 83.1\% of clicks in the visualization mode are content-related, much higher than 57.6\% in the list mode. The visualization \textbf{encourages more interactions with the content}, as the graph visualization simplifies the management operations with an intuitive representation. 
Finally, the visualization and list view have approximately equal chances to lose users to the other mode during usage sessions. 

\begin{table}
\caption{Statistics through log analysis.}
\begin{center}
\begin{tabular}{lcc}
\hline
 Mode & List & Visualization\\
\hline
Number of sessions & 960 & 1,192 \\
Time per session (sec) & 197.4 & 172.3 \\ 
Clicks per session & 12.4 & 17.3 \\
Content-related clicks & 57.6\% & 83.1\%\\
Switch to other mode & 32.3\% & 33.4\% \\
\hline
\end{tabular}
\end{center}
\label{table:log}
\vspace{-1em}
\end{table}


\section{Conclusion and Future Work}

To complement the weaknesses of current social tagging tools, we designed and implemented a new social bookmark manager. It visualizes the tripartite relationships in a more accessible and friendly display, and incorporates richer contextual information. Our design goals include facilitating annotation management with visualization, providing contextual maps, and highlighting social relationships while avoiding information overwhelming. A case study on a common scenario and the system log analysis suggest that these goals have been achieved by our design and implementation. \emph{GalViz} smoothens the process of examining the contextual information and adding new resources; at the same time, users are encouraged to interact with the system and explore the information more efficiently. The \textit{GalViz} social bookmark manager (\url{GiveALink.org/collection/show}) is part of the GiveALink platform.

We presented a set of preliminary evaluations. To learn more about the limitations of \emph{GalViz} and possible improvements in future versions, a more thorough and complete evaluation should be considered. As part of the future work, we plan to employ Amazon's \emph{Mechanical Turk}, a crowd-sourcing marketplace, to measure the usability of the system with several specifically designed tasks. 
Close user studies, including face-to-face conversations and user behavior observations, are necessary as well for gaining a deeper understanding of user requirements.


\bibliographystyle{abbrv}
\bibliography{bmanager_ref}

\end{document}